 \documentclass[preprint2]{aastex}

\def\Mjup{\hbox{$M_{\rm Jup}$}}
\def\Rjup{\hbox{$R_{\rm Jup}$}}

\def\chisq{\mbox{$\chi^2$}\ }

\slugcomment{Submitted to AJ}

\shorttitle{Ground-based near-IR observations of CoRoT-2b}
\shortauthors{Alonso et al.}

\begin{document}


\title{Ground-based near-IR observations of the secondary eclipse of CoRoT-2b \thanks{Based on observations obtained with the William Herschel Telescope, operated by the Isaac Newton Group of Telescopes and situated on the Observatorio del Roque de los Muchachos of the Instituto de Astrof\'\i sica de Canarias in La Palma, Spain.} }


\author{
R. Alonso\altaffilmark{1},
H.J. Deeg\altaffilmark{2,3},
P. Kabath\altaffilmark{4,5},
M. Rabus\altaffilmark{2,3,6}}


\altaffiltext{1}{Observatoire de Gen\`eve, Universit\'e de Gen\`eve, 51 Ch. des Maillettes, 1290 Sauverny, Switzerland}
\altaffiltext{2}{Instituto de Astrof\'\i sica de Canarias, 38205 La Laguna, Spain}
\altaffiltext{3}{Departamento de Astrof\'\i sica, Universidad de La Laguna, 38205 La Laguna, Spain}
\altaffiltext{4}{Institute of Planetary Research, DLR, Rutherfordstr. 2, 12489 Berlin, Germany}
\altaffiltext{5}{European Southern Observatory, Alonso de C\'{o}rdova 3107, Casilla 19001, Santiago de Chile, Chile}
\altaffiltext{6}{Present address Departamento de Astonom\'\i a y Astrof\'\i sica, Pontificia Universidad Cat\'olica de Chile, Casilla 306, Santiago 22, Chile}


\begin{abstract}
We present the results of a ground-based search for the secondary eclipse of the 3.3~\Mjup \,transiting planet CoRoT-2b. We performed near infrared photometry using the LIRIS instrument on the 4.2~m William Herschel Telescope, in the $H$ and $K_s$ filters. We monitored the star around two expected secondary eclipses in two nights under very good observing conditions. For the depth of the secondary eclipse we find  in $H$-band a 3~$\sigma$ upper limit of 0.17\%, whereas we detected a tentative eclipse with a depth of 0.16$\pm$0.09\% in the $K_s$-band. These depths can be translated into brightness temperatures of T$_H<$2250~K and T$_{K_s}$ = 1890$^{+260}_{-350}$~K, which indicate an inefficient re-distribution of the incident stellar flux from the planet's dayside to its nightside. Our results are in agreement with the CoRoT optical measurement \citep{alo09} and with Spitzer 4.5 and 8 $\mu$m results \citep{gill09b}. 
\end{abstract}





\section{Introduction}

\label{sec:intro}
The observation of a secondary eclipse, or occultation of an exoplanet by its host star, is one of the few techniques that allows to extract information about the surface or atmospheric conditions of these bodies. In the infrared, it has already been possible to detect the thermal emission of several Hot Jupiters (starting from \citealt{char05,dem05}), and reveal some features of their atmospheres, such as thermal inversions at high atmospheric altitudes in planets that receive very strong fluxes from their host star (\citealt{knu08,knu09}). These studies were only possible using the Spitzer and HST space telescopes. Using new observing techniques or new instrumentations, the precisions of ground-based photometric time-series have recently shown a significant improvement(\citealt{alo08b,john09,gill08,sou09a}). Very recently, the Na absorption on one exoplanet was detected from the ground \citep{redfield}, and three independent teams announced ground-based detections of the secondary eclipses of different exoplanets (\citealt{sing09,mooij09,gill09}), in $z$ and $K_s$ bands, thus extending the studies of exoplanets' atmospheres to ground facilities. The near infrared windows close to 2 $\mu$m allow to explore a wavelength domain that falls between the detections in the optical accessible now to space missions such as CoRoT \citep{baglin} or Kepler \citep{borucki}, and the wavelengths above 3.5 $\mu$m in the reach of Spitzer, for which \cite{gill09b} recently reported measurements on CoRoT-2b. Thus, high precision near infrared measurements have the potential to provide additional constraints to the models of the atmospheres of these objects.

Orbiting an active G star with a 1.74~d period, the massive planet (3.3\Mjup) CoRoT-2b \citep{alo08a} shows an intriguingly large radius of 1.47\Rjup. While several of the Hot Jupiters show this anomalous and currently poorly understood bloating, CoRoT-2b poses additional challenges to most of the proposed mechanisms to provide extra sources of energy (e.g. \citealt{bod01,show,winn,burr}) due to its large mass. For instance, invoking increased opacities or kinetic energy transport in the planetary interior fails for CoRoT-2b \citep{guillot}. According to these authors, the models point towards a young age for the planet/star system, and tidal dissipation due to either circularization or synchronization of the planet. The effect of on-going tidal circularization might be revealed as a shift of the secondary eclipse time with respect to its expected position according to a circular orbit.

Due to its short period and its large radius, CoRoT-2b is among the exoplanets with the largest expected secondary eclipse depth, and should thus be detectable from ground based facilities. This fact, and the unsolved issues with the radius of the planet, motivated us to perform the observations described in this work.

After describing the way the observations were performed in the next Section, we present the data analysis  in detail in Section 3, which led to an upper limit for the secondary eclipse depth in $H$-filter and a tentative detection of the secondary eclipse in $K_s$. Section 4 will be devoted to the discussion of our results and their implications.

\section{Observations}
\label{sec:obs}

\begin{figure}[!th]
\epsscale{0.9}
\plotone{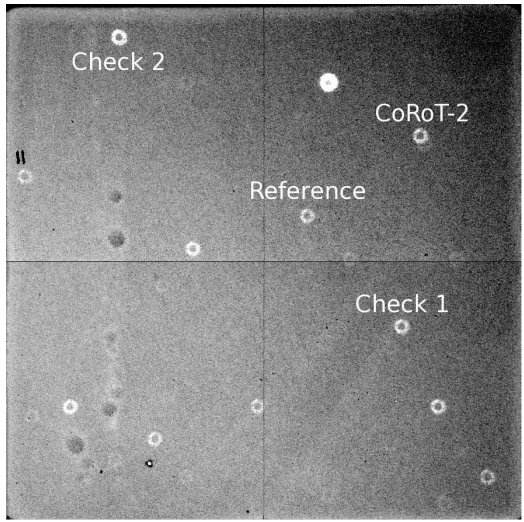}
\caption{A sample raw image of the observed field in the $K_s$ filter, showing the location of the target, the reference star used, and the two check stars that were used in Sect.~\ref{sec:sec_ks} to estimate the effect of the systematic noises.} 
\label{fig:figfind}
\end{figure}

\begin{figure}[!th]
\epsscale{0.9}
\plotone{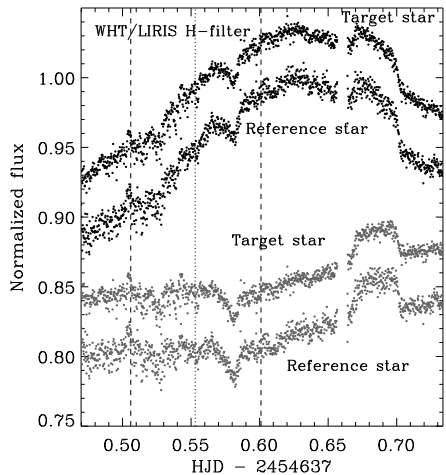}
\caption{Light curves of CoRoT-2b and the reference star in the $H$ filter, before correcting for the changing airmass (top, black points) and after correction (bottom, grey points). Vertical lines show the expected moments of eclipse ingress, center and egress. See Section~\ref{sec:sec_h}. } 
\label{fig:fig4}
\end{figure}

We observed CoRoT-2b during the two nights of 12th and 19th June 2008, using the LIRIS near-IR imager and spectrograph \citep{acosta} on the 4.2~m William Herschel Telescope on La Palma island. The observations were scheduled around the new moon, and the weather conditions were very good, with a seeing varying smoothly between 0.45\arcsec\, and 0.95\arcsec\, on the night of June 12th, and between 0.6\arcsec\, and 1.6\arcsec\, on June 19th. We used the image mode of the LIRIS instrument, with a $K_s$ filter (1.99--2.31$\mu$m) on the first night, and a $H$ filter (1.47--1.78$\mu$m) on the second one. In order to avoid saturation of the detector and to minimize the effect of detector cosmetics, we used the same technique as described for observations with the 1.5~m Carlos S\'anchez Telescope in \cite{alo08b}, that is, we severely defocussed the telescope, and obtained a ring-shaped PSF with an outer radius of about  24~pixels (6\arcsec). The exposure time was 15~s on the first night and 17~s on the second one, with a typical overhead to read the detector and store the data of about 3~s. The flux on individual pixels of the target's PSF was never above 15K ADUs, a dynamical regime where the detector is known to have a good linearity. Therefore, we avoided the necessity to perform non-linearity corrections of the data as done by \cite{mooij09}. In total, we obtained 1380 useful frames on the night of June 12th, and 1096 on June 19th, observing in both nights for about 7.5~h around the expected times of secondary eclipses.
A sample $K_s$ filter image is presented in Figure~\ref{fig:figfind}, showing the position of the target and the reference and check stars that were used during the analysis.

Following again \cite{alo08b}, we chose not to use any dithering pattern, but instead took twice each night a series of 10 images with the star shifted by 30\arcsec\, (\emph{offset images}), in order to monitor the evolution of the image background, keeping the same exposure time. We also obtained detector calibration frames following standard techniques in IR photometry: before sunset, series of 30 images of the dome with the dome-light `on' and `off' were used to build a master flat-field image, by subtracting a combination of the images taken with the lights `on' to the combination of the images with the lights `off'. Unfortunately, we obtained noisier results when we attempted to correct our images by a division of this flat field, and we thus chose not to perform flat-fielding corrections. 

\section{Analysis}
\label{sec:ana}

\begin{figure}[!th]
\epsscale{1}
\plotone{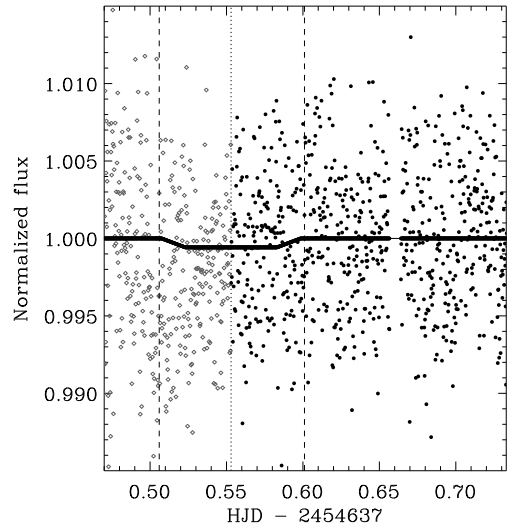}
\caption{Fully corrected curve of CoRoT-2b in the $H$ filter. Vertical lines show the expected moments of eclipse ingress, center and egress. The solid line shows the best-fit trapezoid to the second-half of the secondary eclipse (points in black) . The data points before the center of the eclipse (grey diamonds) are suspicious of uncorrected instrumental effects and have not been considered for the eclipse depth determination. See full description in Section~\ref{sec:sec_h}.} 
\label{fig:fig6}
\end{figure}

\begin{figure}[!th]
\epsscale{0.9}
\plotone{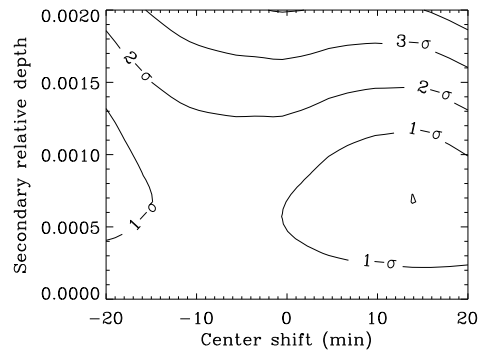}
\caption{The \chisq space for different eclipse center times and depths in the $H$-filter dataset showed in Figure~\ref{fig:fig6}, indicating the 1,2 and 3$\sigma$ confidence limits, using the third method described in Section~\ref{sec:sec_h}. We conservatively interpret these results as a 3-$\sigma$ upper limit of 0.17\% in this wavelength.} 
\label{fig:fig8}
\end{figure}

In order to search for the signal of the secondary eclipse in the two data sets, we adapted the IDL routines used in \cite{alo08b}. The \emph{offset images} of each night were median-combined. We subtracted the mean sky background level from this combined \emph{offset image}, after measuring it in a region without stars. The remaining image $I_{\mathrm{offset}}$ will thus account partially for the hot and dead pixels in the image, as well as for the effects from uncorrected cosmetics in the image. We obtained the best results (lowest dispersion in the final light curve) when we subtracted $I_{\mathrm{offset}}$ from each science image, without performing any other standard calibration. Cosmic ray impacts and remaining hot pixels were detected by applying the IDL's implementation of the top-hat morphological operator (e.g. Serra 1983), which works by applying an opening (dilation) operation to the image, and then subtracting the result from the original image. This way, small features with high flux levels in the image can be identified; they were replaced by a linear interpolation to adjacent pixels. Finally, from each science image we subtracted a sky background level that had been measured in a region without apparent stars.

\begin{figure}[!th]
\epsscale{0.9}
\plotone{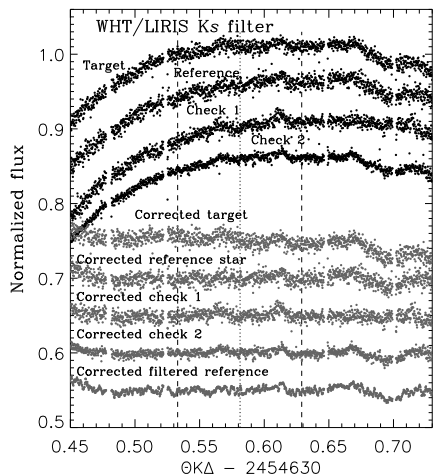}
\caption{Similar to Figure~\ref{fig:fig4}, giving the light curves for the target and reference stars in the $K_s$ filter, before (top, black dots)  and after correcting for the changing airmass (bottom, grey dots). Two check stars at similar distances to the reference were also measured to validate the analysis method, and to measure the stability of the reference star. See Section~\ref{sec:sec_ks}.} 
\label{fig:fig1}
\end{figure}

The centroid of the target star was calculated by fitting a Gaussian with a non-zero revolution axis to the ring-shaped PSF. Since the target has a close companion at 4\arcsec\ distance  with overlapping PSFs (1.72 mags fainter in $K_s$ and 1.8~mags fainter in $H$, from the 2MASS catalogue) that causes the centroid not to be well centered on the target, we added systematic shifts (of -0.7 and 2.7~pixels in the $x$ and $y$ coordinates of the detector) to the calculated target centroid. We also obtained photometry from the ring-shaped psf's of several reference stars with similar fluxes as the target. To avoid the introduction of additional noise from individual fits to the centers of the reference stars, we applied fixed position offsets between the target and reference star apertures. The flux was then summed inside a circular aperture centered on each star. We tested apertures between 24 and 32~pixels in radius, and obtained the best results (less dispersion in the final curve) when using 32~pixels for the $H$ filter data and 26~pixels for the $K_s$ filter night, due to the slightly different defocus applied on the two observing nights. The residual background levels were measured in a ring around each star (between 2 and 2.5 times the aperture radius). The mean level of the target's flux was 7.8$\times10^6$ $e^-$ on the night of the $K_s$ filter observations, and 1.5$\times10^7$ $e^-$ on the night of the $H$ filter data. We recorded for each data point the residual sky background levels measured in the ring around each star, the initially subtracted sky background level, the $x$ and $y$ positions of the stars on the detector, and the airmass, in order to be able to study the possible correlations of the flux curves against those parameters. In the following analysis though, only the values of the positions of the stars on the detector were used to justify the exclusion of the first data points obtained in the $H$-filter data.

\subsection{Analysis of the $H$-filter light curve}
\label{sec:sec_h}

We corrected the effect of the airmass by fitting a cubic polynomial to the off-eclipse sections of the light curve until the star passed the meridian around 3h UT (or 0.62 HJD in Figure~\ref{fig:fig4}), and applied the fitted polynomial to the whole curve. The same procedure was used for the reference star, which was chosen among 8 different stars as the one whose curve had a similar shape as the target. We noticed that even in the relatively small field of view of the instrument, the stars that were further from the target exhibited different behavior as the target, making them useless for building a reference star from a combination of several stars. The initial and corrected fluxes of the target and the reference star are plotted in Figure~\ref{fig:fig4}. 

\begin{figure}[!th]
\epsscale{0.9}
\plotone{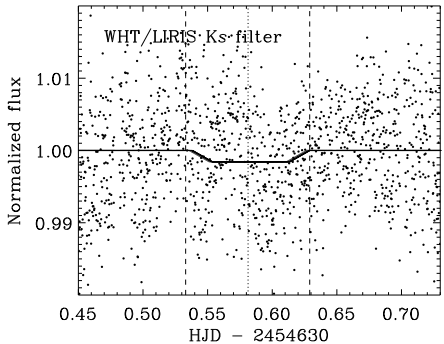}
\caption{Light curve of CoRoT-2b in the $K_s$ filter around the phase of secondary eclipse. The vertical lines show the expected times of first contact, eclipse center and fourth contact, according to the ephemeris in \cite{alo08a}. The solid line plots the best fit model when the eclipse center time is fixed to ephemeris, and the dashed line the best fit model when the center time is also left free (superposed in this plot). See Section~\ref{sec:sec_ks}.} 
\label{fig:fig3}
\end{figure}

Next, we corrected for the flux of the close companion to the target star, which provided an extra 19\% to the target flux in the $H$-band. After the division of the target's flux by the reference star, there still remained some low frequency noise at a level of few 0.001 that we corrected by fitting a 4$^{th}$-order polynomial to the parts outside of the predicted secondary eclipse. The final light curve is presented in Figure~\ref{fig:fig6}. Looking at Figure~\ref{fig:fig4}, the dispersion of the data is clearly higher from the beginning of the observations until approximately the mid-eclipse moment, coinciding with the observations at bigger airmass (starting from 1.8) and more jitter in the guiding of the telescope, which is most notable in the y target-positions in our data (the standard deviation of the recorded y-values improves from 0.9 pixels at the beginning of the night to 0.3 pixels at the end). We suspect that this bigger jitter is due to tensions in the telescope mount when pointing at high airmass. We thus discarded these points in the further analysis, because they are suspicious of being affected by uncorrected instrumental or atmospheric effects (grey points in Figure~\ref{fig:fig6}). Their inclusion in the analysis would result in a deeper and temporally more severely shifted eclipse than the one presented here.

We applied three different methods to evaluate the depth and significance of the secondary eclipse signal. In the first method, we:
\begin{itemize}
\item removed the best fit model from the data\footnote{A simplified model with only depth, center, total duration and ingress/egress duration of the occultation as parameters.}, where the only fitted parameter was the depth, whereas the total duration and ingress/egress duration were fixed to the values of the primary eclipse,
\item shifted circularly the residuals: in order to keep the red noise of the data unaffected, all the time stamps $t_i$ were re-assigned to data points located at a randomly selected distance $R$ in number of points $t_{i+R}$; when $i+R>N$, then the re-assigned time stamp was $t_{i+R-N}$, where $N$ is the total number of points,
\item reinserted the best fit model,
\item re-evaluated the fitted depth.
\end{itemize}
After several hundred realizations (with different randomly selected values of $R$), the final depth, which takes into account the effect of red noise in the data, is 0.06$\pm$0.03\%. We plot this model as the solid line in Figure~\ref{fig:fig6}. 

The second method consisted of the fitting of two Gaussians to 1) the flux-distribution of points inside the full eclipse (138 points in the $H$-filter data between HJD - 2454637 = 0.553 and 0.585, thus avoiding the moments of expected egress time in the case of a circular orbit) and 2) a subset of the points outside the eclipse (there are a total of 532 points) with the same number of points as in 1),  and comparing their centers. This fit was performed to 500 subsets of 2), with randomly chosen starting points and widths  (ranging from 0.0001 to 0.0007 in normalized flux values) of the histogram bins used in the construction of the flux distribution. The resulting depth is of 0.02$\pm$0.06\%. 

Finally, we explored the \chisq distribution of residuals from a fit to a model in a grid of eclipse centers (from -20 to +20 minutes from the expected center-time) and depths (from 0.01\% to 0.2\%) of the secondary eclipse. The minimum \chisq and the 1, 2 and 3-$\sigma$ confidence levels are presented in Figure~\ref{fig:fig8}, and the best fitted depth using this method is 0.07$\pm$0.05\%. None of the three methods provided a statistically significant detection of the secondary eclipse in the $H$-filter, and we can only safely provide a 3-$\sigma$ upper limit for the secondary eclipse depth of 0.17\%.

\subsection{Analysis of the $K_s$ filter light curve}
\label{sec:sec_ks}

\begin{figure}[!th]
\epsscale{0.9}
\plotone{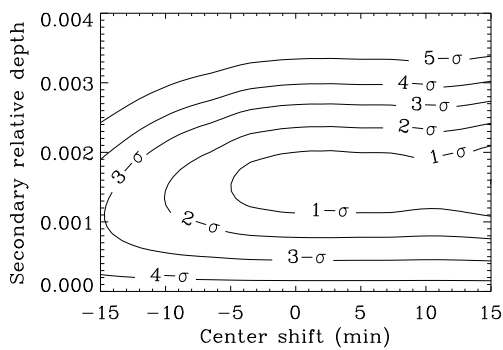}
\caption{The \chisq space for different centers and depths of the secondary eclipse in the $K_s$-filter (Figure~\ref{fig:fig3}), and the 1,2,3,4,5$\sigma$ confidence limits. A secondary eclipse with a depth of 0.16$\pm$0.05\% is tentatively detected at the expected time for the case of a circular orbit. See Section~\ref{sec:sec_ks}.} 
\label{fig:figchisq}
\end{figure}

\begin{figure}[!th]
\epsscale{0.9}
\plottwo{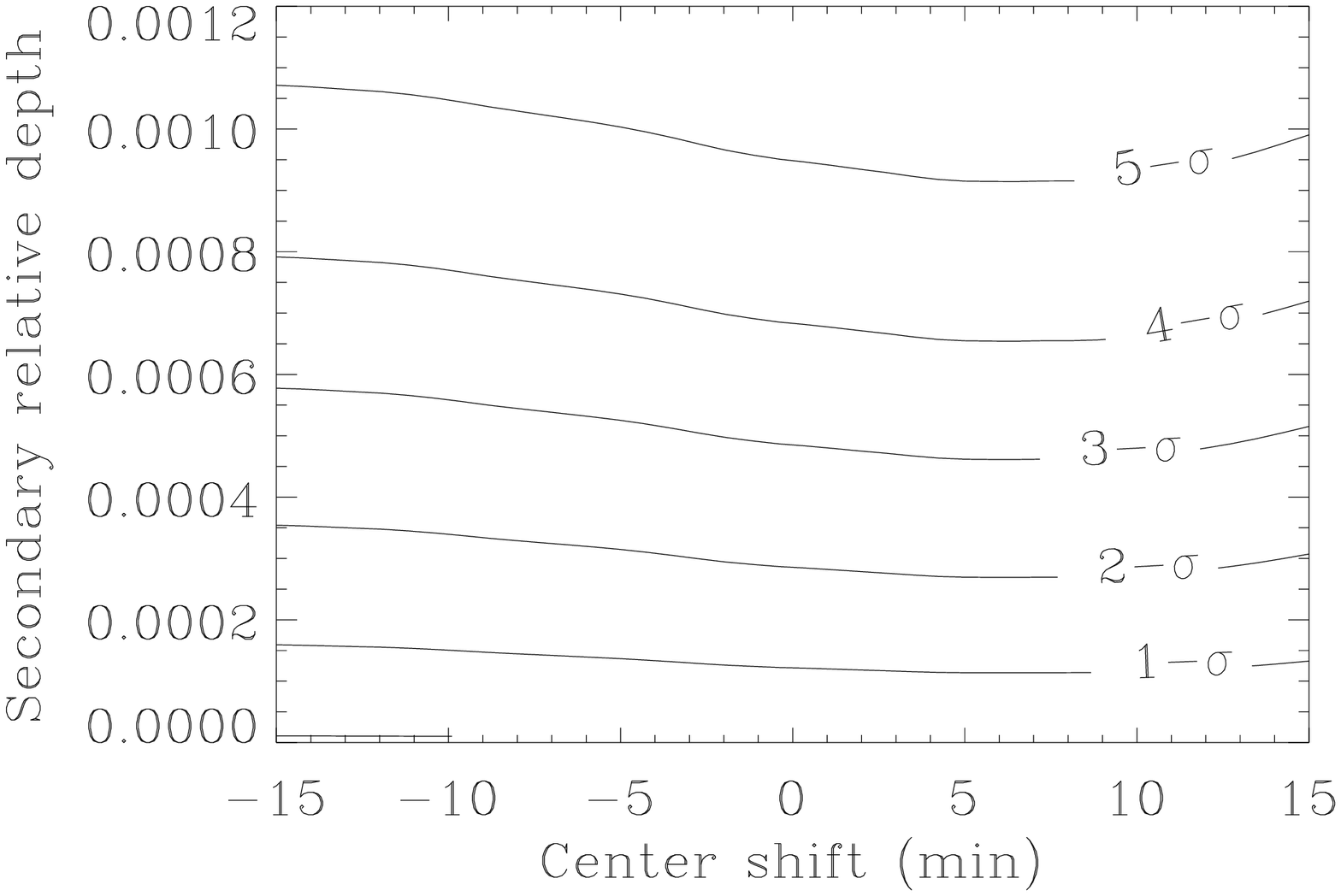}{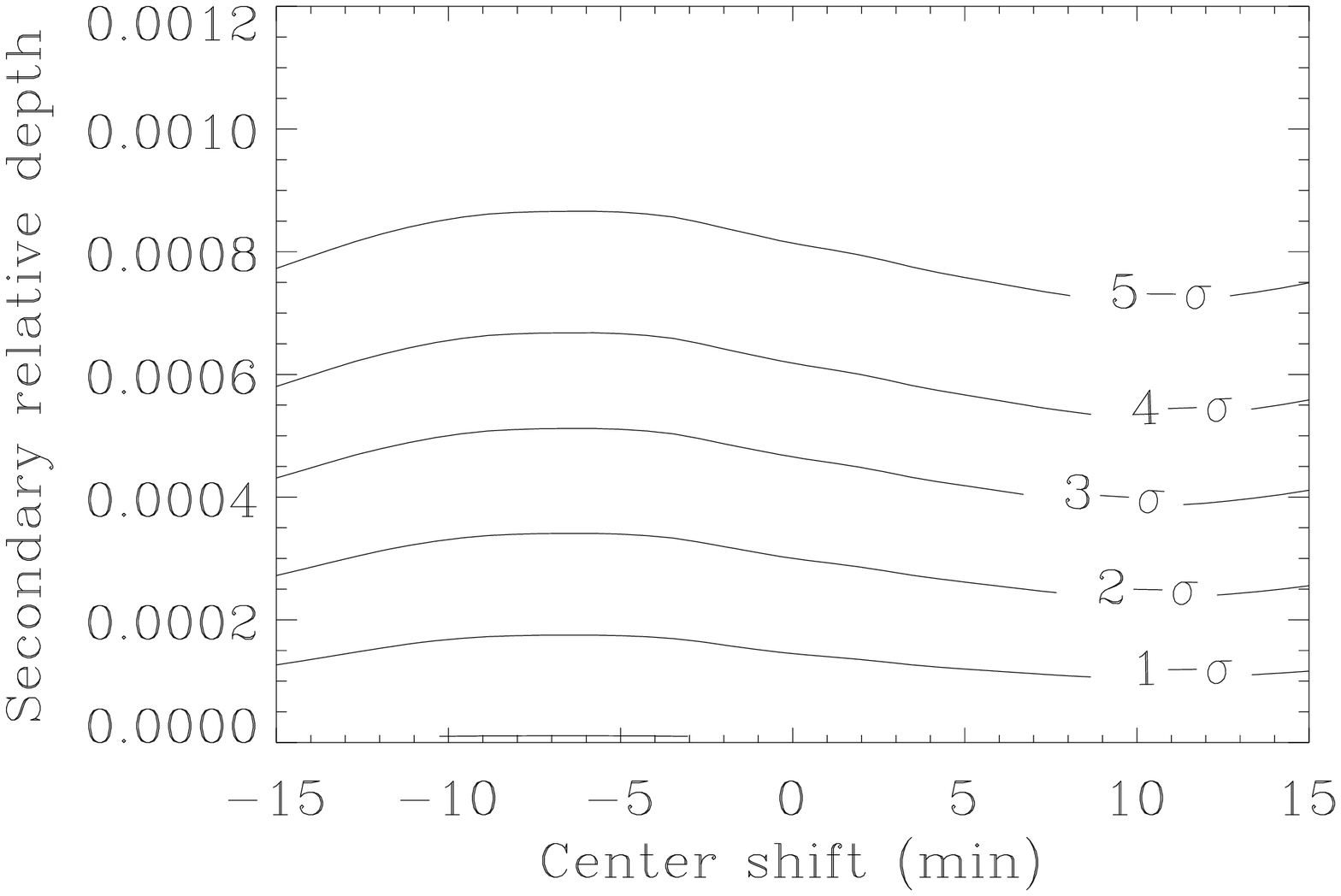}
\caption{The same analysis of Figure~\ref{fig:figchisq} performed on the two check stars. No signal is detected with 3-$\sigma$ upper limits of 0.06 and 0.05\%, thus reassuring the detection presented in the previous Figure.} 
\label{fig:figchisq2}
\end{figure}

We analyzed the $K_s$-filter light curve in a very similar way as the $H$-filter one. We chose an aperture with a radius of 26 pix, and we again corrected for the flux of the close companion to the target star, which provided an extra 16\% of the flux. The plots before and after correction for the airmass with a low order polynomial (in this case a line provided good enough results) are shown in Figure~\ref{fig:fig1}. A careful inspection of that Figure reveals that there are slight differences in the behaviors of the stars, for instance just after 0.67, that depend on the apparent distance in the sky among the different stars. In order to verify that the signal we are searching for is not due to this residual systematic noise, we repeated the same analysis with two check stars located at different parts of the detector and at opposite directions from the reference star. To reduce the high frequency noise of the reference star, we applied a 4-points median filter. The final light curve (after division by the flux of the filtered reference star) shows a tentative secondary eclipse event (Figure~\ref{fig:fig3}). We evaluated the depth and significance of the secondary eclipse using the same three methods described in the $H$-filter section. Model fitting and shifting of residuals (method 1) resulted in a measured depth of 0.17$\pm$0.09\%. The fit to different gaussians (method 2, with 246 points considered to be completely inside eclipse: $0.56<$HJD - 2454630$<0.61$, and 801 points in the out of transit sample) gave a depth of 0.15$\pm$0.15\%, while the \chisq mapping (plotted in Figure~\ref{fig:figchisq}) resulted in an eclipse with a depth of 0.16$\pm$0.05\%. The different error bars reveal the presence of systematic noise in the data that are intrinsically taken into account in methods 1) and 2), but not in the \chisq mapping. We chose as a final value 0.16$\pm$0.09\% for the eclipse depth measurement, i.e., the mean value of the three methods. The \chisq mapping was also performed on the two check stars, and the Figure~\ref{fig:figchisq2} shows 3-$\sigma$ upper limits of 0.06 and 0.05\% respectively for a secondary eclipse, re-assuring the detection of the secondary eclipse in the $K_s$-band.

\section{Discussion}

Our analysis of the near infrared observations obtained at the $H$ and $K_s$ bands has allowed to establish a 3-$\sigma$ upper limit to the occultation of CoRoT-2b in the $H$-band of 0.17\%, while the $K_s$ filter observations reveal a secondary eclipse of 0.16$\pm$0.09\%.

In order to convert the detected depths (or upper limits) in the two band passes into brightness temperatures, we took a modeled spectrum of a G8V star from the \cite{pickles} library. We calibrated the model in order to have the same total flux as a 5625~K Planck function. Under the assumption of a black-body planetary emission, and zero Bond albedo, we computed the flux of the planet for different temperatures. Using the $H $ and $K_s$ filter response functions\footnote{\tiny{http:$//$www.ing.iac.es$/$Astronomy$/$instruments$/$liris$/$liris\_configuration.html}}, we could then estimate the secondary eclipse depth for different temperatures. The 3-$\sigma$ upper limit on the $H$-filter corresponds to an upper limit of the brightness temperature T$_H<$2250~K, while the $K_s$-band measurement implies a black-body emission of T$_{K_s}$=1890$^{+260}_{-350}$~K. If we assume that the planet is in thermal equilibrium, then the measured re-distribution factor is of $f$=$0.57^{+0.39}_{-0.32}$, which points to an inefficient re-distribution of the incident stellar flux from the planet's dayside to its nightside. These results are in agreement, despite their bigger uncertainties, with the values obtained from the analysis of the optical light curve from the CoRoT satellite \citep{alo09}, of T$_{\mathrm{CoRoT}}$=1910$^{+90}_{-100}$K. Recently, \cite{gill09b} published secondary eclipse depths from Spitzer observations, of 0.510$\pm$0.042\% and 0.41$\pm$0.11\% at 4.5 and 8$\mu$m respectively, which led these investigators to also favor models of inefficient heat-distribution.

CoRoT-2b falls in the pM-class of planets according to the classification by \cite{fort}, which exhibit hot stratospheres (temperature inversions), caused by currently debated or unknown upper atmosphere absorbers. According to these authors, such an effect might lead to the detection of ``anomalously" deep secondary eclipses in the near infrared. In the previous interpretation, we have neglected the reflected light component, which is expected to be smaller than the thermal emission at these wavelengths.

Additional observations will help to provide a firm detection in the $H$ band, as well as to improve and confirm the observed $K_s$ band depth of the secondary eclipse. 

\acknowledgments

We thank the staff at the Observatorio del Roque de los Muchachos for help during the observations. We are grateful to M. Santander for support using LIRIS. We want to thank the referee for the very useful comments that helped to improve the manuscript. 
R.A acknowledges support by the grant CNES-COROT-070879 and the IAC grant ``Programa de Acceso a Grandes Instalaciones Cient\'\i ficas". P.K. also acknowledges financial support provided by this program and by DLR. R.A., H.J.D. and M.R. acknowledge support by grant
ESP2007-65480-C02-02 of the Spanish Science and Innovation Ministry.



{\it Facilities:} \facility{WHT (LIRIS)}.

\end{document}